\documentclass[9pt,twocolumn,twoside]{osajnl}
\usepackage{graphicx}
\usepackage{amsmath,amssymb}
\usepackage{graphicx}
\usepackage{color}
\usepackage{bm}
\usepackage{float}
\journal{ol} 

\setboolean{shortarticle}{true}

\title{Pulse reverse-engineering for strong field-matter interaction}

\author[1,2]{Du Ran}
\author[1]{Bin Zhang}
\author[3]{Ye-Hong Chen}
\author[4,*]{Zhi-Cheng Shi}
\author[4,$\dagger$]{Yan Xia}
\author[1,5]{Reuven Ianconescu }
\author[1]{Jacob Scheuer}
\author[1,$\ddag$]{Avraham Gover}

\affil[1]{Department of Electrical Engineering Physical Electronics, Tel Aviv University, Ramat Aviv 69978, Israel}
\affil[2]{School of Electronic Information Engineering, Yangtze Normal University, Chongqing 408100, China}
\affil[3]{Theoretical Quantum Physics Laboratory, RIKEN Cluster for Pioneering Research, Wako-shi, Saitama 351-0198, Japan}
\affil[4]{Fujian Key Laboratory of Quantum Information and Quantum Optics (Fuzhou University), Fuzhou 350116, China}
\affil[5]{Shenkar College of Engineering and Design 12, Anna Frank St., Ramat Gan, Israel}

\affil[*]{Corresponding author: szc2014@yeah.net}
\affil[$\dagger$]{Corresponding author: xia-208@163.com}
\affil[$\ddag$]{Corresponding author: gover@eng.tau.ac.il}

\begin{abstract}
We propose a scheme to control the evolution of a two-level quantum system in the strong coupling regime based on the idea of reverse-engineering.
A coherent control field is designed to drive both closed and open two-level quantum systems along user predefined evolution trajectory without utilizing the rotating-wave approximation (RWA).
As concrete examples, we show that complete population inversion,  an equally weighted coherent superposition, and even oscillation-like dynamics can be achieved.
As there are no limitations on the coupling strength between the control field and matter, the scheme is attractive for applications such as accelerating desired system dynamics and fast quantum information processing.

\end{abstract}

\setboolean{displaycopyright}{true}

\begin{document}

\maketitle

Manipulating physical systems with the electromagnetic field is the cornerstone for the investigation of light-matter interactions. \cite{VAd2017,GOa2019,AVr2020}.
Especially, designation of reliable time-dependent control field to accurately control a quantum system is the key for quantum information processing.
Various recent proposed methods are devoted to this topic, such as adiabatic control \cite{JRk1989,XDt2015}, shortcut-to-adiabaticity (STA) \cite{CXi2010,CYh2014,GOd2019}, optimal control \cite{NKh2005,SHw2019}, Lyapunov control \cite{Shu2008,SCh2012}, etc.
However, most of these studies are based on the RWA which is appropriate only in the weak coupling regime, resulting in long manipulation time.

Recent technological advances have made it possible to realize strong coupling strength between field and matter \cite{KOc2019,FOr2019}. 
For instance, studies on superconducting systems \cite{ATS2004,Mac2018}, optomechanical systems \cite{DMa2016}, semiconducting systems \cite{YSo2016}, Bose-Einstein condensates \cite{SHo2007}, and nitrogen-vacancy centers (NV) \cite{FUc2009} have shown that strong or even ultra-strong couplings can be achieved.
Therefore, it is now possible to engineer systems with strong driving fields to accelerate manipulation precess.
Whereas, such a coupling regime may limit the application of RWA, e.g., in a NV system, Liu \textit{et al.}  \cite{XLi2014} have shown that RWA is broken when the frequency of field of $10\times2\pi$ $GHz$ and the coupling strength reaches $2\pi$ $GHz$.
However, this regime has been in the focus of theoretical and experimental interest on fundamental grounds \cite{JKb2000,SAs2007}.
Thus the schemes based on the RWA should be reconsidered and modified.
Indeed, a variety of ground-breaking techniques for pulse design have been formulated without the RWA.
For instance, the population transition with STA has been reconsidered by Chen \textit{et al.}  \cite{JCh2015}  and  Ib\'{a}\~{n}ez \textit{et al.}  \cite{SIb2015}. 
Fast and accurate qubit operations of a single spin with optimal control have been demonstrated experimentally  \cite{JSc2014}.
While, most of the pulse-design schemes are typically formulated to drive the interested system into a desired final state, sometimes along specific instantaneous eigenstate \cite{YHc2016,ABa2016,FPe2018,LKy2019,HLm2018,DRa2017}.
It is of great interest to control not only the final state superposition but also the exact path, allowing for monitor system dynamics at any moment of time during the interaction with the field.

Recently, it is shown by Golubev \textit{et al.}  \cite{NVg2014,NVg2015}, Medina  \textit{et al.} \cite{IMd2019}, Csehi \cite{ACs2019} and  Ran \textit{et al.} \cite{DRa2020} that within the RWA, a two-level  quantum system can be driven from arbitrary initial state into a desired final state along user-prescribed evolution trajectory based on the idea of reverse-engineering.  
In this letter, we propose a scheme to achieve a desired trajectory in the strong coupling regime where the RWA maybe invalid.
The coherent control field is related to the particular trajectory on the Bloch sphere.
As examples for using our approach, we employ the designed control field for population inversion, superposition state generation, and decay-like Rabi oscillation, which demonstrate the powerful control ability of the scheme.
As the RWA  is not employed,  the scheme is suitable for fast manipulation of a two-level quantum system as desired with strong field.
Thus, it offers opportunities to test advanced concepts of quantum control and achieve fast state operation compared to previous approaches.


We consider a two-level quantum system with ground state $|g\rangle$ and excited state $|e\rangle$ interacting with a laser electric field $E(t)=E_0(t)\cos[\varphi(t)]$, where $E_0(t)$ is the amplitude and $\varphi(t)$ contains the frequency $\omega_L$ and phase $\phi(t)$, i.e., $\varphi(t)=\omega_Lt+\phi(t)$.
In the Schr\"{o}dinger picture, the Hamiltonian of the system in the electric dipole approximation without using the RWA can be written as ($\hbar=1$) \cite{IBa2011} 
 \begin{eqnarray}\label{HS}
\hat{H}_s(t) = \frac{1}{2}[\omega_0(t) \hat{\sigma}_z + \Omega_R(t)(\sigma_+ +\sigma_-) (e^{i\varphi(t)} + e^{-i\varphi(t)})],
\end{eqnarray}
where $\omega_0(t)$ is the transition frequency of the two-level system, which depends on time, e.g., controlled by Stark shifts. 
$\Omega_R(t)$ is the time-dependent Rabi frequency, assumed to be real without loss of generality. 
$\hat{\sigma}_j$ $(j=x, y, z)$ are the pseudospin operators, namely Pauli matrices. 
$\hat{\sigma}_{\pm}=(\hat{\sigma}_{x} \pm i\hat{\sigma}_{y})/2$ are the the raising or lowering operators.
By defining $H_{\varphi}(t)=\dot{\varphi}(t)\hat{\sigma}_z/2$, the exact Hamiltonian in a field-adapted interaction picture is given by 
 \begin{eqnarray}\label{HI}
\hat{H}(t) =\hat{U}^{\dagger}_{\varphi}(\hat{H}_s - \hat{H}_{\varphi})\hat{U}_{\varphi}
=\frac{1}{2}\left
(
\begin{tabular}{cccc}
$-\Delta(t)$      &    $\Omega(t)$ \\
$\Omega^*(t)$   &   $\Delta(t)$  \\
   \end{tabular}
\right
),
\end{eqnarray}
where $\hat{U}_{\varphi} = \exp\{-i\int^t_0H_{\varphi}(t')\}dt'$ is the unitary operator of the transition, $\Omega(t)=\Omega_R(t)[1+\exp\{-2i\varphi(t)\}]$,  and  $\Delta(t)=\omega_0(t) -\dot{\varphi}(t)$.

Here, we consider the general case that the two-level system interacts with the environment which induces thermal noise and dephasing with rates $\Gamma$ and $\gamma$, respectively. 
In such scenario, the dynamics of the system is given by  \cite{HJc1999}
$\dot{\hat{\rho}}(t) =-i[\hat{H}(t),\hat{\rho}(t)] + \frac{\gamma}{2}D_{de}[\hat{\rho}(t)] + \Gamma D_{th}[\hat{\rho}(t)],$
where $D_{de}[\hat{\rho}(t)]=\hat{\sigma}_z\rho(t)\hat{\sigma}_z-\hat{\rho}(t)$ and $D_{th}[\hat{\rho}(t)]=\bar{n}[2\hat{\sigma}_+\hat{\rho}(t)\hat{\sigma}_- - \{\hat{\sigma}_-\hat{\sigma}_+,\hat{\rho}(t)\}] + (\bar{n}+1)[2\hat{\sigma}_-\rho(t)\hat{\sigma}_+ - \{\hat{\sigma}_+\hat{\sigma}_-,\hat{\rho}(t)\}]$ with $\bar{n}$ being the effective photon number.
Based on the optical Bloch equation $\dot{\hat{\rho}}(t)=\frac{1}{2}[\hat{\mathbb{I}} + \dot{u}(t)\hat{\sigma}_x +\dot{v}(t)\hat{\sigma}_y +\dot{w}(t)\hat{\sigma}_z]$, with the idea of reverse engineering \cite{DRa2020}, $\varphi(t)$ and 
$\Omega_R(t)$ for the coherent control field  are straightforwardly calculated 
\begin{subequations}\label{CFS}
\begin{eqnarray}
&&\varphi(t)=\int [\omega_0(t) - \Delta(t)]dt,\\
&&\Omega_R(t) = \Omega(t)/ [1+\cos(2\varphi)],
\end{eqnarray}
\end{subequations}
where $\hat{\mathbb{I}}$ is the identity operator, $h(t)=\textrm{Tr}[\hat{\sigma}_h\hat{\rho}(t)]$ $(h = u, v ,w)$, and
\begin{subequations}\label{fOD}
\begin{eqnarray}
&&\Omega(t) = \frac{2\Gamma}{v(t)}[1+w(t) +2\bar{n}w(t)] +\frac{\dot{w}(t)}{v(t)},      \\
&&\Delta(t) = \frac{\tilde{\Gamma}u(t) + \dot{u}(t)}{v(t)}.
\end{eqnarray}
\end{subequations}

Let us proceed to illustrate this method by some concrete examples.
We first consider the closed quantum system case, i.e., $\gamma=\Gamma=0$.
Then, according to Eq. (\ref{fOD}), one can see that $\Omega(t)=\dot{w}(t)/v(t)$ and  $\Delta(t)=\dot{u}(t)/v(t)$.
It is clear that $u^2(t)+v^2(t)+w^2(t)=1$ for the general evolution of a closed two-level quantum system.
Therefore, after providing two of the concrete form of the predefined functions $u(t)$, $v(t)$ and $w(t)$,  the other one that predefines the evolution of the two-level system is automatically defined.
For instance, a given $u(t)$ and $w(t)$, then $v(t)=\sqrt{1-u^2(t)-w^2(t)}$. 

For population transition, we assume that the function $w(t)$ has the following form: $w(t)=a_i(1-g(t)) + a_fg(t)$,
where $g(t)=1/(1+\exp(-\alpha t))$ goes smoothly from 0 to 1 with a real and positive parameter $\alpha$ controlling the time duration of transition from initial state to final state.
Note that the function $w(t)$ is related to the evolution of population $P_{g(e)}=\langle g(e)|\hat{\rho}(t)|g(e)\rangle$.
Here, we choose the function $u(t)$ to have the Gaussian envelope: $u(t)=Ae^{-\frac{1}{2}{(t-\tau)^2}/{\sigma^2}}$, where the real parameter $A$ is the height of the curve's peak, $\tau$ is the position of the center of the peak, and $\sigma$ controls the width of the ``bell''.

\begin{figure}[htpb]
\centering
\scalebox{0.22}{\includegraphics{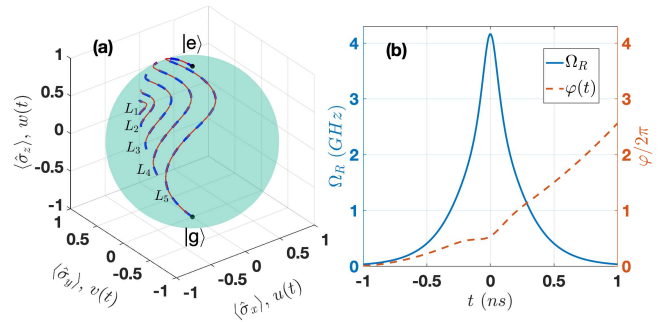}}
\caption{(a) Different expected trajectories (red-solid lines) and the corresponding numerical simulated  trajectories (blue-dash lines) on the Bloch sphere defined by $\{u(t)$, $v(t)$, $w(t)\}$ and $\{\langle\hat{\sigma}_x\rangle, \langle\hat{\sigma}_y\rangle, \langle\hat{\sigma}_z\rangle\}$, respectively. 
The same parameters for all lines are: $\alpha=0.01\times10^{12}$, $\tau=0$, $\sigma=100\times10^{-12}$, $\Gamma=\gamma=0$, and $\omega_0(t)$ increases linearly with time from 1 $GHz$ to 15 $GHz$ ,
while for $L_1$: $a_i=-0.1$, $a_f=0.1$,  $A=0.1$;
for $L_2$: $a_i=-0.25$, $a_f=0.25$,  $A=0.2$
for $L_3$: $a_i=-0.5$, $a_f=0.5$,  $A=0.4$;  
for $L_4$: $a_i=-75$, $a_f=0.75$,  $A=0.6$;  
and for $L_5$: $a_i=-1$, $a_f=1$,  $A=0.8$. 
(b) The Rabi frequency $\Omega_R$ and phase $\varphi$ of the coherent field for obtaining the trajectory $L_5$.
}\label{Blo}
\end{figure}

Because the interaction pictures lead to the same Schr\"{o}dinger dynamics and same populations, next, we perform the simulations in the  interaction picture.  
To exhibit that the scheme can drive the system from an arbitrary initial state to a desired final state, in Fig. \ref{Blo}(a), we show different trajectories on the Bloch sphere by choosing different $a_i$ and $a_f$ and adjusting the parameters in the functions $u(t)$ and $w(t)$.
From the simulation, one can see that each prescribed trajectory are obtained exactly.
By choosing a particular trajectory, in Fig. \ref{Blo}(b), we show the corresponding $\Omega_R(t)$ and $\varphi(t)$ linearly increased the transition frequency $\omega_0(t)$.
An observation of Fig. \ref{Blo}(b), one can note that the shape of $\Omega_R(t)$ and $\varphi(t)$ are very simple and the maximum modulus of the Rabi frequency $|\Omega_R|_{max}$ have the same order of magnitude of the transition frequency $\omega_0(t)$.
This means that the scheme works well in the strong coupling regime.

\begin{figure}[htpb]
\centering
\scalebox{0.23}{\includegraphics{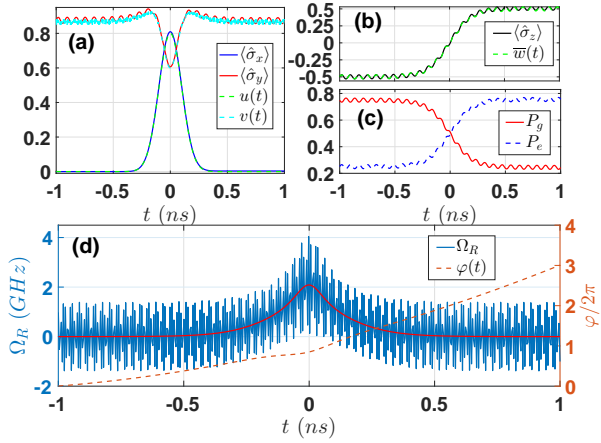}}
\caption{(a) The predefined functions $u(t)$, $v(t)$, $\langle\hat{\sigma}_x\rangle$ and $\langle\hat{\sigma}_y\rangle$  as a function of time. 
(b) The predefined functions $w(t)$ and $\langle\hat{\sigma}_z\rangle$ as a function of time. 
(c) The evolution of populations $P_g$ and $P_e$. 
(d) The time evolutions of $\Omega_R$ and $\varphi$ of the coherent control field. The red line is the time evolutions of $\Omega_R$ in the smooth varying case ($\chi=0$).
The parameters for the simulation are: $a_i=-0.5$, $a_f=0.5$, $\chi=0.03$, $\Omega=0.08\times10^{12}$, and the remaining parameters are the same with Fig. \ref{Blo}.
}\label{ooo}
\end{figure}

In cases where the RWA  dose not apply, it has been widely demonstrated that the transition process is always oscillatory due to the existence of the counter-rotating terms \cite{IBa2011,YHk2017}.
Obviously, the scheme here can also present such oscillatory evolution process.
For instance, by defining the function $\overline{w}(t)=w(t)+\chi\cos(\Omega t)$, we can obtain oscillatory population evolution, where $\chi$ adjusts the amplitude and $\Omega$ controls the periodicity  of oscillation.
In Fig. \ref{ooo}, a typical oscillatory dynamics is presented. 
Figure \ref{ooo}(a) shows the predefined functions $u(t)$ and $v(t)$ together with the evolution of the numerical expectation values $\langle\hat{\sigma}_x\rangle$ and $\langle\hat{\sigma}_y\rangle$.
The analytical function $\overline{w}(t)$ and numerical expectation value $\langle\hat{\sigma}_z\rangle$ are shown in Fig. \ref{ooo}(b).
One can note that the numerical results of $\langle\hat{\sigma}_j\rangle$ ($j= x, y, z$) coincide with the analytical expected oscillation evolution.  
Figure \ref{ooo}(c) is the populations evolution of $P_e$ and $P_g$.
Such a dynamics is driven by the control field with the corresponding $\Omega_R$ and $\varphi(t)$ given in Fig. \ref{ooo}(d).
For the oscillatory dynamics,  the Rabi frequency $\Omega_R$ oscillate with the period and amplitude determined by $\Omega$ and $\chi$, respectively.

\begin{figure}[htpb]
\centering
\scalebox{0.228}{\includegraphics{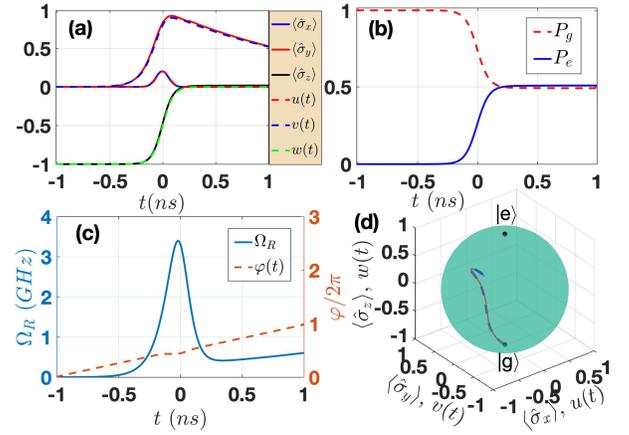}}
\caption{
(a) The predefined functions $h(t)$ ($h=u, v ,w$) and $\langle\hat{\sigma}_j\rangle$ ($j= x, y, z$) as a function of time.
(b) The evolution of populations $P_g$ and $P_e$. 
(c) $\Omega_R$ and $\varphi$ as a function of time. 
(d) The expected  trajectory (red-solid lines) and the corresponding numerical simulated  trajectory (blue-dash lines) on the Bloch sphere defined by $\{u(t)$, $v(t)$, $w(t)\}$ and $\{\langle\hat{\sigma}_x\rangle, \langle\hat{\sigma}_y\rangle, \langle\hat{\sigma}_z\rangle\}$, respectively.
The parameters for the simulations are: $a_i=-1$, $a_f=0$, $\alpha=0.02\times10^{12}$, $\omega_0=5$ $GHz$, $A=0.2$, $\tau=0$, $\sigma=60\times10^{-12}$, $\bar{n}=0$, $\Gamma=0.1$ $GHz$, and $\gamma=1$ $GHz$.}\label{Opens}
\end{figure}

Generally, the dynamics of the system will unavoidably be influenced by the environment.  
In the following, we study the reverse-engineering scheme in the open quantum system by taking into account the presence of dephasing and thermal noise. One illustrative example is depicted in Fig. \ref{Opens}.
In the simulation, the initial state of the system is in the ground state $|g\rangle$ and the final state is the superposition state $|\psi(+\infty)\rangle=\frac{1}{\sqrt{2}}(|g\rangle+|e\rangle)$ corresponding to $a_i=-1$ and $a_f=0$, respectively.
Figure \ref{Opens}(a) is the analytical predefined functions $h(t)$ ($h=u, v ,w$) and the numerical results of 
$\langle\hat{\sigma}_j\rangle$ ($j= x, y, z$).
Figure \ref{Opens}(b) depicts the evolution of  populations  $P_g = |\langle g|\hat{\rho}(t)|g\rangle|$ and $P_e = |\langle e|\hat{\rho}(t)|e\rangle|$. 
Figure \ref{Opens}(c) shows the Rabi frequency $\Omega_R$ and $\varphi$ that relate to the coherent control field. 
From the simulation, it is clear that the desired dynamics is obtained because the analytical predefined functions are perfectly coincide with the numerical simulations, which is confirmed by the trajectory on the Bloch sphere in Fig. \ref{Opens}(d).
We note that the populations can be maintained with the non-vanishing control field even in the presence of dissipation.
Hoever, the coherence, defined as $|\langle\hat{\sigma}_x\rangle-i\langle\hat{\sigma}_y\rangle|/2$, decay with time due to the dissipation effect, resulting in the trajectory moving inside the Bloch sphere.

\begin{figure}[htpb]
\centering
\scalebox{0.22}{\includegraphics{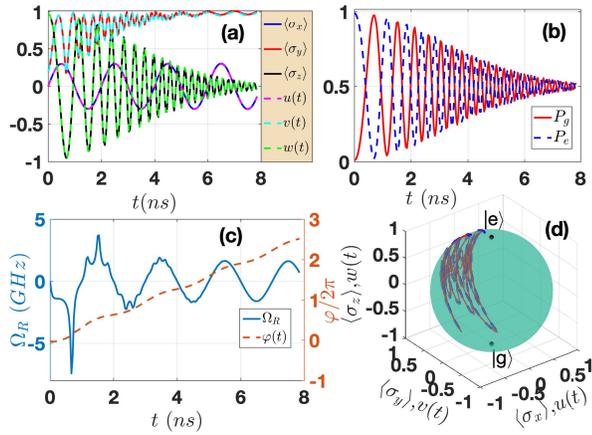}}
\caption{
(a) The predefined functions $h(t)$ ($h=u, v ,w$) and $\langle\hat{\sigma}_j(t)\rangle$ ($j= x, y, z$) as a function of time.
(b) The evolution populations $P_g$ and $P_e$. 
(c) The $\Omega_R$ and $\varphi$ as a function of time. 
(d) The expected trajectory (red-solid lines) and the corresponding numerical simulated  trajectory (blue-dash lines) on the Bloch sphere defined by $\{u(t)$, $v(t)$, $w(t)\}$ and $\{\langle\hat{\sigma}_x\rangle, \langle\hat{\sigma}_y\rangle, \langle\hat{\sigma}_z\rangle\}$, respectively.
The parameters for the simulations are: $b=2\times10^{18},$ $a=0.05\times10^{18}$, $k_1=0.98$, $k_2=0.3$, $\Omega_1=\Omega_2=\pi$ $GHz$,  and $\bar{n}=\gamma=\Gamma=0$.
}\label{PC2}
\end{figure}

We have shown that the scheme is feasible for population transfer in both closed and open quantum systems.
Next, as an additional example, we show that the scheme can also be used to obtain arbitrary decay of Rabi oscillations \cite{KNz2015,VVd2009}.
The predefined functions $w(t)$ and $u(t)$ are chosen as:  $w(t)= k_1 e^{-at^2} \cos(\Omega_1t + bt^2), u(t)=k_2 \sin(\Omega_2t)$,
where $0\leqslant k_1\leqslant1$ and $0\leqslant k_2\leqslant1$ are used for adjusting the amplitude of $w(t)$ and $v(t)$, respectively. 
$\Omega_1$ and $\Omega_2$ are used for controlling the ocsillation frequency of the population and coherence.   
$a$ is a constant which can be used for controlling the population decay rate. $b$ is a real chirp parameter.
The common Rabi oscillation process is obtained by setting the parameters $a=b=0$ where $k_1$ controls the amplitude of Rabi oscillation.
One simulation result is given in Fig. \ref{PC2}.
The analytical predefined functions $h(t)$ ($h=u, v ,w$) and the numerical results of $\langle\hat{\sigma}_j\rangle$ ($j= x, y, z$) are presented in Fig. \ref{PC2}(a), which are also exhibited on the Bloch sphere in Fig. \ref{PC2}(d).
Figure \ref{PC2}(b) depicts the evolution of  populations $P_g$ and $P_e$. 
Figure \ref{PC2}(c) shows $\Omega_R(t)$ and $\varphi(t)$ as a function of time. 
The simulation results evidence that arbitrary decay of Rabi oscillations of the system can be obtained in the strong coupling regime.

Numerous merits of the scheme are emerged over previous ones. 
First, by using the density representation instead of the wave function for the system \cite{NVg2014,NVg2015}, it allows us to control both closed and open quantum systems as desired.
By taking into account the presence of environmental noise and arbitrary initial states, the scheme may have applications for controlling molecular or condensed-matter systems \cite{AIk2011}.
Since there is no limitation on the phase of coherence where the previous schemes assume constant \cite{NVg2014,NVg2015,IMd2019} , as a result, not only the populations but also the coherence can be controlled.
In addition, previous schemes \cite{NVg2014,NVg2015,IMd2019,ACs2019,DRa2020} require resonant interaction between light-matter interaction for controlling the evolution trajectory.
Here, such a requirement is not needed, allowing the control of the system for time-varying transition frequency.
Most importantly, the scheme allows us to control the system in the strong or even ultra-strong coupling regime by adjusting the parameters in the predefined functions, which may be applied to ultafast quantum system control \cite{CSe2019}.
For instance, for population inversion, by choosing 15 orders of magnitude for $\alpha$ and $\sigma$, the magnitude of the control field would increases by factor 3 and the manipulation time decrease to the order of picosecond.
This solves the problem of limitation by increasing $\alpha$ to shorten the transition time, as has been indicated in Ref. \cite{DRa2020}. 
While, it has been indicated that the expected trajectory is obtained only when the RWA is well satisfied \cite{DRa2020}, i.e., $\omega_0/|\Omega_R|_{max}>20$.
Besides, the scheme here can be regarded as the generalization of proposals given by Zlatanov and Vitanov \cite{KNz2020}, in which the adiabatic condition is considered.

In summary, we have proposed a simple scheme to design coherent field for controlling arbitrarily  the evolution of two-level systems in the strong-coupling regime.
The flexibility of the scheme ensures that arbitrary desired dynamics can be obtained, such as population inversion, superposition state generation and oscillation-like dynamics synthesis.
Importantly, due to the field obtained without utilizing the RWA, fast manipulation of system dynamics is allowed by adjusting the parameters in the predefined functions.
Besides, the pulse reverse-engineering scheme maybe applied to realize universal quantum logic gates.
Thus, it may have potential interest to a variety of physical problems involving qubits driven by an external field, e.g., in quantum metrology \cite{PAI2013} and quantum simulations \cite{AFr2008}.

\textbf{Funding.} 
This work was supported by the National Natural Science Foundation of China under Grants No. 11805036, No. 11575045, and No. 11674060. 
Y.-H.C. is supported by the Japan Society for the Promotion of Science (JSPS) Foreign Postdoctoral Fellowship No.~P19028.

\textbf{Disclosures.} The authors declare no conflicts of interest.

\newcommand{\noopsort[1]}{} \newcommand{\singleletter}[1]{#1}



\begin{thebibliography}{99}
\newcommand{\enquote}[1]{``#1''}


\bibitem{VAd2017} D. Valente, F. Brito, and T. Werlang, Dynamic Stark shift induced by a single photon, Opt. lett. 42, 1692 (2017).

\bibitem{GOa2019} A. Gover, R. Ianconescu, A. Friedman, C. Emma, N. Sudar,  P. Musumeci, and C. Pellegrini,  Superradiant and stimulated-superradiant emission of bunched electron beams, Rev. Mod. Phys. 91, 035003 (2019).
\bibitem{AVr2020} A. Gover and A. Yariv, Free-Electron–Bound-Electron Resonant Interaction, Phys. Rev. Lett. 124, 064801 (2020).

 
\bibitem{JRk1989} J. R. Kuklinski, U. Gaubatz, F. T. Hioe, and K. Bergmann, Adiabatic population transfer in a three-level system driven by delayed laser pulses, Phys. Rev. A 40, 6741 (1989).
\bibitem{XDt2015} X. D. Tian, Y. M. Liu, C. L. Cui, and J. H. Wu, Population transfer and quantum entanglement implemented in cold atoms involving two Rydberg states via an adiabatic passage, Phys. Rev. A 92, 063411 (2015).


\bibitem{CXi2010} X. Chen, I. Lizuain, A. Ruschhaupt, D. Gu\'{e}ry-Odelin, and J. G. Muga, Shortcut to adiabatic passage in two-and three-level atoms, Phys. Rev. Lett. 105, 123003 (2010).
\bibitem{CYh2014} Y. H. Chen, Y. Xia, Q. Q. Chen, and J. Song, Efficient shortcuts to adiabatic passage for fast population transfer in multiparticle systems, Phys. Rev.  A 89, 033856 (2014).
\bibitem{GOd2019} D. Gu\'{e}ry-Odelin, A. Ruschhaupt, A. Kiely ,  E. Torrontegui, S. Mart\'{i}nez-Garaot, and J. G. Muga, Shortcuts to adiabaticity: concepts, methods, and applications, Rev. Mod. Phys. 91, 045001 (2019).
 

\bibitem{NKh2005} N. Khanejaa, T. Reissb, C. Kehletb, T. S. Herbr\"{u}ggen, and S. J. Glaser, Optimal control of coupled spin dynamics: design of NMR pulse sequences by gradient ascent algorithms, J. Magn. Reson. 172, 296 (2005).
\bibitem{SHw2019} S. H.Wu, M. Amezcua, and H.Wang, Adiabatic population transfer of dressed spin states with quantum optimal control, Phys. Rev. A 99, 063812 (2019).


\bibitem{Shu2008} S. Kuang and S. Cong, Lyapunov control methods of closed quantum systems, Automatica 44, 98 (2008).
\bibitem{SCh2012} S. C. Hou, M. A. Khan, X. X. Yi, D. Y. Dong, and I. R. Petersen, Optimal Lyapunov-based quantum control for quantum systems, Phys. Rev. A 86, 022321 (2012).


\bibitem{KOc2019} A. F. Kockum, A. Miranowicz, S. De Liberato, S. Savasta, and  F. Nori, Ultrastrong coupling between light and matter. Nature Reviews Physics {1}, 19 (2019). 

\bibitem{FOr2019} P. Forn-D\'{i}az, L. Lamata, E. Rico, J. Kono, and E. Solano, Ultrastrong coupling regimes of light-matter interaction. Rev. Mod. Phys. {91}, 025005 (2019). 


 
\bibitem{ATS2004}A. T. Sornborger, A. N. Cleland, and M. R. Geller, Superconducting phase qubit coupled to a nanomechanical resonator: Beyond the rotating-wave approximation, Phys. Rev. A 70, 052315 (2004).

\bibitem{Mac2018} V. Macr\`{i}, F. Nori, and A. F. Kockum, Simple preparation of Bell and Greenberger-Horne-Zeilinger states using ultrastrong-coupling circuit QED. Phys. Rev. A, 98, 062327 (2018).



\bibitem{DMa2016} D. Malz and A. Nunnenkamp, Optomechanical dual-beam backaction-evading measurement beyond the rotating-wave approximation, Phys. Rev. A 94, 053820 (2016).

\bibitem{YSo2016} Y. Song, J. P. Kestner, X. Wang, and S. D. Sarma, Fast control of semiconductor qubits beyond the rotating-wave approximation, Phys. Rev. A 94, 012321 (2016).

\bibitem{SHo2007} S. Hofferberth, B. Fischer, T. Schumm, J. Schmiedmayer, and I. Lesanovsky, Ultracold atoms in radio-frequency dressed potentials beyond the rotating-wave approximation, Phys. Rev. A 76, 013401 (2007).

\bibitem{FUc2009} G. D. Fuchs, V. V. Dobrovitski, D. M. Toyli, F. J. Heremans, and D. D. Awschalom, Gigahertz dynamics of a strongly driven single quantum spin, Science 326, 1520 (2009).



\bibitem{XLi2014} X. Liu, G. Y. Fang, Q. H. Liao, and S. T. Liu, Fast multiqubit phase gate in circuit QED beyond the rotating-wave approximation, Phys. Rev. A 90, 062330 (2014).

 
\bibitem{JKb2000} J . K. Boyd, Probability amplitude dynamics for a two-level system, J. Math. Phys. 41, 4330 (2000).
\bibitem{SAs2007} S. Ashhab, J. R. Johansson, A. M. Zagoskin, F. Nori, Two-level systems driven by large-amplitude fields, Phys. Rev. A 75, 063414 (2007).


\bibitem{JCh2015} J. Chen and L. F. Wei, Implementation speed of deterministic population passages compared to that of Rabi pulses, Phys. Rev. A 91, 023405 (2015).
\bibitem{SIb2015} S. Ib\'{a}\~{n}ez, Y. C. Li, X. Chen, and J. G. Muga, Pulse design without the rotating-wave approximation, Phys. Rev. A 92, 062136 (2015).

\bibitem{JSc2014} J. Scheuer, X. Kong, R. S. Said, J. Chen, A. Kurz, L. Marseglia, J. F Du, P. R Hemmer, S. Montangero , T.  Calarco, B. Naydenov, and F Jelezko, Precise qubit control beyond the rotating wave approximation,  New J.  Phys. 16, 093022 (2014).




\bibitem{YHc2016} Y. H. Chen, Y. Xia, Q. C. Wu, B. H. Huang, and J. Song, Method for constructing shortcuts to adiabaticity by a substitute of counterdiabatic driving terms, Phys. Rev. A 93, 052109 (2016).
\bibitem{ABa2016} A. Baksic, H. Ribeiro, and A. A. Clerk, Speeding up adiabatic quantum state transfer by using dressed states, Phys. Rev. Lett. 116, 230503 (2016).



\bibitem{FPe2018} F. Petiziol, B. Dive, F. Mintert, and S.Wimberger, Fast adiabatic evolution by oscillating initial Hamiltonians, Phys. Rev. A 98, 043436 (2018).
\bibitem{LKy2019}  K. Y. Liao, X. H. Liu, Z. Li, and Y. X. Du, Geometric Rydberg quantum gate with shortcuts to adiabaticity,  Opt. Lett. 44, 4801 (2019).

\bibitem{HLm2018} H. L. Mortensen, J. J. W. S\o rensen, K. M\o lmer, and J. F. Sherson, Fast state transfer in a $\Lambda$-system: a shortcut-to-adiabaticity approach to robust and resource optimized control, New J. Phys. 20, 025009 (2018).
\bibitem{DRa2017} D. Ran, Z. C. Shi, J. Song, and Y. Xia, Speeding up adiabatic passage by adding Lyapunov control, Phys. Rev. A 96, 033803 (2017).

\bibitem{NVg2014} N. V. Golubev and A. I. Kuleff, Control of populations of two-level systems by a single resonant laser pulse, Phys. Rev. A 90, 035401 (2014).
\bibitem{NVg2015} N. V. Golubev and A. I. Kuleff, Control of charge migration in molecules by ultrashort laser pulses, Phys. Rev. A 91, 051401(R) (2015).
\bibitem{IMd2019}I. Medina and F. L. Semi\~{a}o, Pulse engineering for population control under dephasing and dissipation, Phys. Rev. A 100, 012103 (2019).
\bibitem{ACs2019} A. Csehi, Control of the populations and phases of two-level quantum systems by a single frequency-chirped laser pulse, J. Phys. B: At. Mol. Opt. Phys.  52, 195004 (2019).

\bibitem{DRa2020} D. Ran, W. J Shan, Z. C Shi, Z. B Yang, J. Song, and Y Xia, Pulse reverse engineering for controlling two-level quantum systems, Phys. Rev. A 101, 023822 (2020)






\bibitem{IBa2011} S. Ib\'{a}\~{n}ez, A. P. Conde, D. Gu\'{e}ry-Odelin, and  J. G.  Muga, Phys. Rev. A 84, 013428 (2011); S. Ib\'{a}\~{n}ez, Y. C. Li, X. Chen, and J. G. Muga, Interaction of strongly chirped pulses with two-level atoms, Phys. Rev. A 92, 062136 (2015). 


\bibitem{HJc1999} H. J. Carmichael, StatisticalMethods in Quantum Optics 1,Master Equations and Fokker-Planck Equations (Springer-Verlag, Berlin, 1999).


\bibitem{YHk2017} Y. H. Kang, Y. H. Chen, B. H. Huang, J. Song, and Y. Xia,   Invariant‐Based Pulse Design for Three‐Level Systems Without the Rotating‐Wave Approximation. Ann. Phys. (Berlin) 529, 1700154 (2017).





\bibitem{KNz2015} K. N. Zlatanov, G. S. Vasilev, P. A. Ivanov, and N. V. Vitanov, Exact solution of the Bloch equations for the nonresonant exponential model in the presence of dephasing, Phys. Rev. A 92, 043404 (2015).
 
 \bibitem {VVd2009} V. V. Dobrovitski, A. E. Feiguin, R. Hanson, and D. D.  Awschalom,  Decay of Rabi oscillations by dipolar-coupled dynamical spin environments, Phys. Revi. Lett. 102, 237601 (2009).
 
 \bibitem {AIk2011} A. I. Kuleff and L. S. Cederbaum, Radiation generated by the ultrafast migration of a positive charge following the ionization of a molecular system, Phys. Rev. Lett. 106, 053001 (2011).


 \bibitem  {CSe2019}  A. Csehi, M. Kowalewski, G. J. Hal\'{a}sz, and \'{A}. Vib\'{o}k,   Ultrafast dynamics in the vicinity of quantum light-induced conical intersections. New J. Phys. {21}, 093040 (2019).
 
\bibitem{KNz2020} K. N. Zlatanov, and  N. V. Vitanov, Adiabatic generation of arbitrary coherent superpositions of two quantum states: Exact and approximate solutions, Phys. Rev. A 96, 013415 (2017); K. N. Zlatanov, and  N. V. Vitanov, Generation of arbitrary qubit states by adiabatic evolution split by a phase jump, Phys. Rev. A 101, 013426 (2020).

\bibitem{PAI2013} P. A. Ivanov and D. Porras, Adiabatic quantum metrology with strongly correlated quantum optical systems, Phys. Rev. A 88, 023803 (2013).

\bibitem{AFr2008} A. Friedenauer, H. Schmitz, J. T. Glueckert, D. Porras, and T. Schaetz, Simulating a quantum magnet with trapped ions,  Nat. Phys. 4, 757 (2008).







\end{thebibliography}
\end{document}